\documentclass[aps,prd,11pt,nofootinbib]{revtex4-2}

\usepackage[utf8]{inputenc}
\usepackage{fancyvrb}
\usepackage{footnote}
\usepackage{braket}
\usepackage{amssymb}
\usepackage{amsmath}
\usepackage[mathscr]{eucal}
\usepackage{epsfig}
\usepackage{textcomp}
\usepackage[T1]{fontenc}
\usepackage{leftindex}
\usepackage{graphicx}
\usepackage{array}
\usepackage{xcolor}
\usepackage{marginnote}

\definecolor{mediumorchid}{rgb}{0.71, 0.32, 0.80}

\def\gm#1#2{{\gamma}_{{#1}}^{{#2}}}

\begin{document}

\title{Quasinormal Modes of Black Holes: Efficient and Highly Accurate Calculations with Recurrence-Based Methods.}

\author{Kristian Benda} 
\email{kribenq@gmail.com}
\author{Jerzy Matyjasek} 
\email{jurek@kft.umcs.lublin.pl}
\affiliation{Institute of Physics,
Maria Curie-Sk\l odowska University\\
pl. Marii Curie-Sk\l odowskiej 1,
20-031 Lublin, Poland}

\begin{abstract}
We discuss new  recurrence-based methods for calculating the complex frequencies 
of the quasinormal modes of black holes. These methods are based on the Frobenius 
series solutions of the differential equation describing the linearized radial 
perturbations. Within the general method, we propose two approaches: the first 
involves calculating the series coefficients, while the second employs 
generalized continued fractions. Moreover, as a consequence of this analysis, we 
present a computationally efficient and convenient method that uses double 
convergence acceleration, consisting of the application of the Wynn algorithm to 
the approximants obtained from the Hill determinants, with the 
Leaver-Nollert-Zhidenko-like tail approximations taken into account. The latter 
is particularly important for stabilizing and enabling the calculations 
of modes with small real parts as well as higher overtones. The method 
demonstrates exceptionally high accuracy. We emphasize that Gaussian elimination 
is unnecessary in all of these calculations. We consider $D$-dimensional 
($3<D<10$) Schwarzschild-Tangherlini black holes as concrete examples. 
Specifically, we calculate the quasinormal modes of the $(2+1)$-dimensional 
acoustic black hole (which is closely related to the five-dimensional 
Schwarzschild-Tangherlini black holes), the electromagnetic-vector modes of the 
six-dimensional black holes and the scalar (gravitational tensor) modes in the 
seven-dimensional case. We believe that the methods presented here are 
applicable beyond the examples shown, also outside the domain of the black hole 
physics. 
\end{abstract}
%\tableofcontents
\maketitle

\section{Introduction.}
\label{sec:secI}

The quasinormal modes of black holes, discovered by Vishveshwara~\cite{vishnu} 
in 1970 and popularized by Chandrasekhar in his influential 
monograph~\cite{chandra}, are the solutions of the perturbation equations 
satisfying the outgoing (i.e., moving away from a potential barrier) boundary 
conditions.  These solutions belong to the discrete set of complex frequencies, 
$\omega$, characterizing oscillations and the rate of damping. Black holes can 
be perturbed in many ways: by external fields, during the process of matter 
accretion, infalling particles, close passage of other astrophysical bodies,
as a result of aspherical gravitational collapse, 
or through collisions with stars, black holes or compact objects. 
Interested reader may consult a few excellent reviews that have been devoted to 
the quasinormal modes and related 
phenomena~\cite{kostas99,berti,konoplya11,nollert2,ppani13}. Especially interesting in 
this regard are the black hole–black hole collisions, such extreme phenomena 
that information about their course reaches the Earth after billions of years in 
the form of the detectable gravitational waves. Roughly speaking, the collision 
process proceeds in the three stages: 
the inspiral, merger and ringdown. The 
signal generated during the first phase informs about the nature of the black 
holes before the collision, whereas the final (ring-down) phase carries information 
about the newly-formed black hole. 

A new generation of gravitational wave detectors is expected to observe the 
ringdown modes in gravitational waveforms, enabling high-precision 
gravitational-wave spectroscopy. (For the status of the fundamental mode 
detection and  the low-lying overtones, see Ref.~\cite{destounis2023} and the 
references therein). It is therefore of paramount importance to have at one's 
disposal not only the accurate characteristics of the quasinormal modes of 
various black hole configurations, but also a deep understanding of how they are 
influenced by surrounding matter and quantum effects. 

At the same time, it has been recognized that certain properties of black holes, 
such as the appearance of the apparent horizon, Hawking radiation, quasinormal 
modes and some optical phenomena can, in principle, be observed in a laboratory. 
As the Hawking effect is probably too weak to be observed directly in the 
astrophysical regime, detection of the created `particles' in a laboratory would 
indicate the universality of this phenomenon. Moreover, in the case of 
the laboratory black holes, we should have substantial control over their specific 
characteristics, which naturally facilitates significantly more rigorous testing 
of the applied computational methods~\cite{visser}.

Therefore there is nothing surprising in the fact that a lot of effort has been 
devoted not only to identifying the quasinormal modes of various black hole 
configurations but also to developing mathematical techniques for their 
determination. Over the years, various approaches to the problem have been 
developed, including direct integration of the perturbation 
equation~\cite{chandra2}, the semi-analytic approach based on the WKB 
expansion~\cite{wkb0,wkb1,konoplyawkb6,jaOP,jago,galtsov,phase1}, the asymptotic 
iteration method~\cite{naylor}, the spectral method~\cite{jansen}, the continued 
fraction method~\cite{leaver,leavergauss,nollert, zhidenko}, and, as we shall 
demonstrate explicitly, closely related to the latter, the Hill determinant 
method~\cite{hill,leavergauss,matyjasek2021accurate}. Although each of the 
aforementioned methods has proven to be a valuable tool for determining 
quasinormal modes, two of them have gained significant popularity: WKB-based 
methods due to their wide applicability and the continued fraction method 
because of its accuracy.

Since the publication of Leaver's seminal paper~\cite{leaver} on the application 
of the continued fraction method for calculating quasinormal modes of black 
holes, it has become, whenever possible, a standard and widely used technique in 
the field. At the same time, its mathematically equivalent counterpart, the Hill 
determinant method, introduced to the black hole physics by Majumdar and 
Panchapakesan~\cite{hill}, has been largely overlooked (see, however, 
Ref.~\cite{leavergauss,matyjasek2021accurate}). The popularity of the Leaver's 
papers should come as no surprise to anyone: all the basic ingredients of the 
continued fraction method are there. Indeed, such notions as the inversions of 
the fractions, convergence and basic implementation of its acceleration and 
stabilization, as well as the simplest one-step Gaussian elimination are studied 
and analyzed. The continued fraction approach received new impetus from the work 
of Nollert~\cite{nollert} and Zhidenko~\cite{zhidenko}, who developed the 
asymptotic methods for evaluating of the so called remainder (tail), i.e., the 
infinite part of the continued fraction that, in the naive approach, is set to 
zero to obtain the $N$-th convergent.

The continued fraction method in its original form however, allows for 
determining quasinormal modes only if the recurrence relation for the 
coefficients of the Frobenius series is of the third-order. To circumvent this 
limitation, while dealing with the higher-order recurrences, the Gaussian 
elimination is routinely used \cite{leavergauss, zhidenko, gauss}. We show that 
this step, although it leads to the correct result, is an unnecessary 
complication. More importantly, it increases computational time and, in a purely 
numerical approach, may introduce additional numerical errors. Contrary to what 
one might initially think, calculating the determinant of the resulting 
tridiagonal matrix analytically may take even more time than computing the 
original one, as its coefficients are rational rather than polynomial functions. 
Moreover, as the Gaussian elimination does not lead to a general, closed-form 
expressions for the new coefficients of the thus constructed three-term 
recurrence~\cite{zhidenko}, it significantly hinders any further analysis of the 
recursion itself. Having all this in mind, we decided to critically revisit the 
methods of calculating the complex frequencies of the quasinormal modes based on 
the recurrences and identify the most efficient and convenient approach that 
utilizes the given Frobenius series solution. It should be emphasized that we 
now live in a quite different `computational reality', where powerful computers 
and sophisticated symbolic algebra packages allowing arbitrary precision 
arithmetic are widely available, enabling (symbolic) calculations of 
unprecedented complexity. 

We start our analyses with the most obvious, although not recommended, approach, 
that consists in the explicit calculations of the coefficients of the Frobenius 
series. Subsequently, we analyze the problem using what is, in a sense, the 
generalization of the inversion of the continued fractions, and, finally, we 
study the applicability of the Hill determinant method.  The latter, in our 
numerical implementation, turns out to be the most efficient. Nevertheless, all 
three methods are just the three faces of the same idea, ultimately leading to 
the equivalent final equations: the only difference lies in the  efficiency of 
adapted computational strategy. In all three cases we take into account the 
appropriate tail terms that considerably improve the quality of the results.  
Neither of the approaches require application of the techniques of the Gaussian 
elimination. And to improve the results thus obtained even further, one can 
accelerate their convergence using some well known techniques, such as the Wynn 
acceleration. To the best of our knowledge most of the results presented here 
are new.

We will use the results described in this paper in the context of the five-, 
six-, and seven-dimensional Schwarzschild-Tangherlini black holes and focus on 
the challenging cases, which undoubtedly include the calculation of $\omega$ for 
higher overtones. These cases should be sufficient to illustrate all the major 
features of the techniques we propose. Moreover, due to the close relationship 
between five- and seven-dimensional black holes, and the (2+1)- and 
(3+1)-dimensional acoustic black holes, respectively, our techniques and results 
may also be helpful for the interpretation of the experimental results. Finally, 
let us observe that a common wisdom regarding overtones is that, due to their 
short lifetime, they are devoid of any astrophysical significance. However, 
recent analyses suggest that their precise understanding is helpful, or even 
essential, in the analysis of signals observed by the gravitational wave 
detectors.

The paper is organized as follows. The master equation describing the linear 
perturbations  of the $D$-dimensional Schwarzschild -Tangerlini is briefly 
discussed in Sec.~\ref{sec:secII}. In Sec.~\ref{sec:secIII} we introduce three 
methods of calculation of the complex frequencies of the quasinormal modes based 
on the recurrence relations. They are in the order of appearance: explicit 
solution of the recurrence equations, the method of the \emph{generalized} 
continued fractions and the Hill determinant method, each of them augmented with 
the asymptotic formula approximating the remainder terms. In 
Sec.~\ref{sec:secIV},  we apply our techniques to a few physically interesting 
cases described by the master equation of the  $D$-dimensional 
Schwarzschild-Tangherlini black holes, with a special emphasis put on its formal 
relations  with the radial equations of the perturbed acoustic black holes. In 
Sec.~\ref{sec:secV} we summarize the obtained results with the emphasis put on 
the hard cases. Finally, in the
appendix, we give  the coefficients of the Nollert-like expansions for five-, 
six-, and seven-dimensional black holes.

\section{The master equation of the $D$-dimensional Schwarzschid-Tangherlini black holes   }
\label{sec:secII}

To illustrate our considerations with a specific example, let us  consider the 
differential (master)  equation describing the radial perturbations of the 
$D$-dimensional Schwarzschild-Tangherlini black hole ($D>3$)~\cite{dim1, dim2, 
rostworowski}
\begin{equation}
\frac {d^2} {d r_{\star}^{2}} \psi + \left\{\omega^{2}  - f(r) \left[ \frac {l(l+D-3)} {r^2}
+ \frac {(D-2)(D-4)} {4r^2} + \frac {(1-j^2)(D-2)^2} {4r^{D-1}} \right]\right\}\psi = 0,
\label{eq:master1}
\end{equation} 
where
\begin{equation}
f(r) = 1 - \frac{1}{r^{D-3}},
\end{equation} 
$j$ characterizes the type of the perturbation: 
\begin{equation}
j=
\begin{cases}
0, &       \text{massless scalar and gravitational tensor perturbations}.  \\ 
2, &       \text{gravitational vector perturbations},\\
\frac{2}{D-2}, &   \text{electromagnetic vector  perturbations}\\
2 -\frac{2}{D-2}, & \text{electromagnetic scalar  perturbations}.\
\end{cases}
\label{eq:typyzab}
\end{equation}
and $r_{\star}$ is the tortoise coordinate defined in a standard way. The 
complex frequencies of the quasinormal modes, $\omega,$ are defined as the 
solutions of this equation that are purely ingoing as $r_{\star} \to -\infty$ 
and purely outgoing as $r_{\star} \to \infty$, with the asymptotic behaviors 
$e^{ -i \omega r_{\star}}$ and $e^{i \omega r_{\star}},$ respectively. An 
interesting observation is the formal resemblance between the master equation 
and the equations governing radial perturbations of other black hole 
configurations. This corresponds to considering in \eqref{eq:master1} values of 
$j$ that are not represented in~\eqref{eq:typyzab}, as well as allowing for $l$ 
to take non-integer values. This means that the techniques we have developed can 
be readily applied in such cases as well, and in what follows, we shall make use of this
opportunity.

The master equation has one irregular singularity at infinity and 
$(D-2)$ regular singular points~\cite{rostworowski}. To construct the 
appropriate Frobenius series, the points on the complex $r$-plane must be 
transformed in such a way that the (transformed) regular singularity located originally at $r = 1$, 
representing the event horizon, becomes the closest point to the (transformed) 
irregular singularity. In doing so, the series has the required radius of 
convergence and the discretization conditions for the quasinormal modes are 
given by the condition of convergence of the series at the infinity. The 
solutions for $D<10$ satisfying quasinormal modes boundary conditions can be 
written as \cite{rostworowski}:
\begin{equation}
 \psi(r) = \left( \frac{r-1}{r} \right)^{-i\omega/(D-3)} e^{i \omega r }
\sum_{k=0}^{\infty} a_{k} \left(\frac{r-1}{r} \right)^{k},
                     \label{eq:expansion_even}
\end{equation}
for the even-dimensional case and
\begin{equation}
 \psi(r) = \left(\frac{r-1}{r+1}\right)^{-i \omega/(D-3)} e^{i \omega r}
\sum_{k=0}^{\infty} a_{k} \left(\frac{r-1}{r} \right)^{k}
                     \label{eq:expansion_odd}
\end{equation}
for the odd-dimensional case and the radius of convergence of the Frobenius 
series is equal to 1. Substitution of the expansions \eqref{eq:expansion_even} 
and \eqref{eq:expansion_odd} into the master equation \eqref{eq:master1} leads 
either to the $(2 D-5)$-term linear homogeneous recurrence relation for the 
expansion coefficients $a_{k}$ in the case of the even-dimensional black hole, or to 
the $(2 D - 6)$-term one in the case of the odd-dimensional black hole. It is worth 
noting that both of these solutions can be modified to obtain a lower-order 
recurrence \cite{analog}. 

\section{Recurrence based methods}
\label{sec:secIII}

To ensure the utmost generality of our analysis, let us consider a 
linear  homogeneous $P$-term recurrence
\begin{equation}
\sum^{P}_{i=1} a_{k+2-i}\gamma_{k}^{i-2}=0, \quad k=0,1,2\ldots,
\label{eq:recd}
\end{equation}
with the initial conditions given by
\begin{equation}
 0=\gamma_{0}^{-1} a_{1} +\gamma_{0}^{0} a_{0}=a_{-1}=a_{-2}=\ldots=a_{2-P},
 \label{eq:reci}
\end{equation}
where $\gamma_{k}^{i-2}$ are polynomial functions of $\omega$ and $k$.

We remind that the homogeneous linear $P$-term recurrence  
has $(P-1)$ fundamental (i.e., linearly independent) solutions \cite{benors}.    
According to the Birkhoff-Trjitzinsky theorem \cite{res}, it is possible to identify 
$P - 1$ distinct asymptotic behaviors, each 
representing a solution that collectively form 
a fundamental system of the recurrence. As 
$\lim_{k \to \infty} |\frac{a_{k}}{a_{k+1}}|=R$ \cite{benors}, 
where $R$ is the radius of 
convergence of the series, we can deduce which solutions are permitted by the 
recursion with the given initial conditions \eqref{eq:reci}. It turns out that
the  asymptotics represented by the following expression (details 
of the calculation are given below): 
\begin{equation}
 a_{k+1}/a_{k} \sim 1-\frac{\sqrt{2 \rho}}{\sqrt{k}}+\ldots,\label{eq:aa} 
 \end{equation}
where $\rho = - i \omega$,
is associated with the solution appropriate for convergence of the Frobenius 
series at infinity (the real part of the second term has to be negative) 
\cite{rostworowski, leaver, zhidenko}, and in this case 
$\lim_{k \rightarrow \infty }a_k = 0$. 
Another allowed behavior is $a'_{k+1}/a'_{k} \sim 
1+\frac{\sqrt{2 \rho}}{\sqrt{k}}+\ldots$, which is associated with the dominant 
solution having divergent coefficients of the series. Finally, we have the 
remaining $(P-3)$  cases of the form $a''_{k+1}/a''_{k} \sim Q + \dots $, where 
$Q$ is some complex number such that $\rvert Q \rvert<1$.

Now, in order to determine the quasinormal modes, we may start, without loss of 
generality, with some value of $a_{0}$, while the remaining coefficients 
can be constructed recursively from the initial conditions, e.g., 
$a_{0}=-\gamma_{0}^{-1}$, $a_{1}=\gamma_{0}^{0}$, and so on. Then, we
calculate $a_{L}$ as a function of the coefficients $\gamma,$  which, in turn, 
depend on $\omega,$ and set $a_{L}$ to zero: 
\begin{equation}
a_{L}(\gamma(\omega))=0 
\label{eq:a}. 
\end{equation} 
This approach is justified by the form of the 
asymptotic behaviors. It turns out that the larger the value of 
$L$
we take, the more higher lying candidates for quasinormal modes can be identified 
as solutions of the resulting equation for $\omega$, while the approximations of the lower lying candidates become increasingly precise.
This equation can be reduced to a polynomial of increasing degree, 
which, in the cases considered in this article, is equal to $2 L + 1$.
Moreover, one expects that the
$L$-th overtone is one of its highest solutions, i.e., the one with the smallest
imaginary part.
However, the approach  based on Eq. \eqref{eq:a}, although simple and natural, is
computationally inefficient due to the appearance of $\gamma_{k}^{-1}$ in the 
denominators of the recursion. Fortunately, an equivalent equation can be 
obtained within the framework of at least two other approaches: the continued 
fraction method \cite{leaver} and the Hill determinant method \cite{hill}.

To study the relationship between these three approaches, we divide Eq.\eqref{eq:recd}
by $a_{k}$ and define 
$r_{k}=\frac{a_{k+1}}{a_{k}}.$  Expressing the result  
solely in terms of the new variable  and subsequently solving the equation with respect to $r_{k}$,
we obtain the upward recurrence
\begin{equation}
r_{k}=\frac{-\gamma^{S}_{k+S}}{\sum_{i=0}^{S} \gamma^{i-1}_{k+S}\prod^{S-i}_{j=1}r_{k+j}}\label{eq:fracrec},
\end{equation}
where  $S=P-2$.
Now, we can start with $r_{0}=-\frac{\gamma_{0}^{0}}{\gamma_{0}^{-1}}$ and 
recursively replace $r_{k+j}$ terms with the right-hand sides of 
Eq.\eqref{eq:fracrec}. This procedure leads to the object that can be referred to as the 
generalized continued fraction. We see that if we want to obtain the equation
equivalent to our initial procedure \eqref{eq:a} we should set the $r_{L'}$ to zero 
$(L'>S)$, and replace $r_{L'-1}, r_{L'-2},\ldots,r_{L'-(S-1)}$ with their values 
obtained from the recurrence \eqref{eq:recd}. Unfortunately, it seems that this method is reasonable only for 
the three-term recurrence, for which it is consistent with Leaver's 
approach~\cite{leaver} and the Pincherle theorem~\cite{gautschi}\footnote{Pincherle 
theorem guarantees that if two independent solutions of a three-term recurrence are in 
dominant-minimal relation in terms of their limiting values, then the quotient 
of consecutive elements of non vanishing minimal solution can be expressed 
as some concrete continued fraction with the recurrence coefficients as its components.}, 
yielding a continued fraction
\begin{equation}
 \gamma^{0}_{0} - \frac{\gamma^{-1}_{0} \,\gamma^{1}_{1}}{\gamma^{0}_{1} -}\, \frac{\gamma^{-1}_{1}\,
 \gamma^{1}_{2}}{\gamma^{0}_{2} -}\, \frac{\gamma^{-1}_{2}\, \gamma^{1}_{3}}{\gamma^{0}_{3} -}\dots = 0.
 \label{cont}
\end{equation}
For the higher-order recurrences we need to calculate $r_{L}$ terms, from 
which we can easily construct  the appropriate algebraic equation and determine the quasinormal modes 
without any additional complications. This can be accomplished by employing equation \eqref{eq:recd} to derive the downward recurrence
\begin{equation}
 r_{k}=-\frac{1}{\gamma^{-1}_{k}}\sum_{i=0}^{S} \gamma^{i}_{k}\prod^{i}_{j=1}\frac{1}{r_{k-j}}
 \label{eq:inver}.
\end{equation}
In this case, for the calculations\footnote{In the actual computational process, it is convenient to set specific 
conditions, such as $\gamma^{i}_{j} = 0$ for $1\leq i \leq S$, $ 0\leq j < i$ and 
$r_{0} = \gamma^{0}_{0}/\gamma^{-1}_{0}$ with $r_{-S}=0$. The expressions $r_{-S+1},\dots ,r_{-1}$ 
can be equated to any non-zero real number.}
we only need to input $r_{0}, r_{1}, r_{2},\ldots,r_{S-1}$. Now, equating
the downward and upward recurrences, one has
\begin{equation}
 \frac{1}{\gamma^{-1}_{k}}\sum_{i=0}^{S} \gamma^{i}_{k}\prod^{i}_{j=1}\frac{1}{r_{k-j}}
 =
 \frac{\gamma^{S}_{k+S}}{\sum_{i=0}^{S} \gamma^{i-1}_{k+S}\prod^{S-i}_{j=1}r_{k+j}}.
 \label{eq:down-up}
\end{equation}
Specializing it to the three-term recurrence, i.e., taking $S=1$, we obtain the continued 
fraction in its $k$-fold inverted form~\cite{leaver}  
\begin{equation}
 \gamma^{0}_{k} -\frac{\gamma^{-1}_{k-1} \,\gamma^{1}_{k}}{\gamma^{0}_{k-1} -}\,\frac{\gamma^{-1}_{k-2} \,
 \gamma^{1}_{k-1}}{\gamma^{0}_{k-2} -}\,\dots -\frac{\gamma^{-1}_{0} \gamma^{1}_{1}}{\gamma^{0}_{0}} = 
\frac{\gamma^{-1}_{k} \,\gamma^{1}_{k+1}}{\gamma^{0}_{k+1} -}\, \frac{\gamma^{-1}_{k+1}\,
 \gamma^{1}_{k+2}}{\gamma^{0}_{k+2} -}\, \frac{\gamma^{-1}_{k+2}\, \gamma^{1}_{k+3}}{\gamma^{0}_{k+3} -}\dots 
 \label{eq:cont}
\end{equation}

For further analysis, let us approach the problem from a different 
perspective. Consider the truncated Hill determinant of the banded matrix of the 
width $S+2,$ where $S$ is the number of subdiagonals, constructed from the 
$\gamma$ 
coefficients~\eqref{eq:recd}
\begin{equation}
\mathcal{H}_{k}=
\begin{vmatrix}
\gamma_{0}^0 & \gamma_{0}^{-1} \\
\gamma_{1}^1 & \gamma_{1}^0 & \gamma_{1}^{-1}\\
\gamma_{2}^2&\gamma_{2}^1 & \gamma_{2}^0 & \gamma_{2}^{-1}\\
\vdots\\
\gamma_{S}^{S} & \ldots&\gamma_{S}^1 & \gamma_{S}^0 & \gamma_{S}^{-1}\\
&\gamma_{S+1}^{S} & \ldots&\gamma_{S+1}^1 & \gamma_{S+1}^0 & \gamma_{S+1}^{-1}\\
&&&&\ddots&\ddots&\ddots\\
&&&\gamma_{k-1}^{S}&\ldots&\gamma_{k-1}^1 & \gamma_{k-1}^0 & \gamma_{k-1}^{-1}\\
&&&&\gamma_{k}^{S}&\ldots&\gamma_{k}^1 & \gamma_{k}^0\\
\end{vmatrix} \label{eq:mat}
\end{equation}
Although the banded matrices are sparse, computation of the determinants of the 
large matrices with the analytic matrix elements may be a real challenge. It 
would therefore be desirable to have a linear recursive formula relating the 
determinant $\mathcal{H}_{k}$ to the determinants of its leading principal 
minors $\mathcal{H}_{i}$ (which are, of course, the Hill determinants of the 
lower order). Iteratively expanding the $k$-th determinant $\mathcal{H}_{k}$ and 
its arising minors along the last column, after some algebra, we get\footnote{A 
structurally similar formula was recently derived for the matrix valued Hill 
determinant in Ref.~\cite{matrix}. }
\begin{equation}
\mathcal{H}_{k}=\sum^{S}_{m=0}(-1)^m \gamma_{k}^{m} \mathcal{H}_{k-(m+1)}\prod^m_{j=1}\gamma_{k-j}^{-1}, 
\label{eq:hillrec}
\end{equation}
where the initial conditions are $\mathcal{H}_{0}=\gamma_{0}^0$, $\mathcal{H}_{-1}=1$, and 
all $\mathcal{H}_{-k}$ for $k\geq 2$ vanish. This elegant formula is useful 
for both theoretical analysis and numerical calculations. 

Depending on our goals, equation~\eqref{eq:hillrec} can be transformed either 
into an upward recurrence or a downward recurrence. Indeed, defining 
$R_{k}=\frac{\mathcal{H}_{k}}{\mathcal{H}_{k-1}}\frac{1}{\gamma_{k}^{-1}}$ we 
arrive at the desired resuls. For example, for the upward recurrence, one has
\begin{equation}
R_{k}=\frac{\gamma^{S}_{k+S}}{\sum_{i=0}^{S} (-1)^{i-P} \gamma^{i-1}_{k+S}\prod^{S-i}_{j=1}R_{k+j}}.
\label{eq:remainderrec}
\end{equation}
Zhidenko~\cite{zhidenko} derived an equivalent formula directly from the recurrence \eqref{eq:recd}. 
It should be noted that setting $R_{k}=-r_{k}$ in this equation yields \eqref{eq:fracrec}.
Proceeding in a similar manner in the case of the downward recurrence, we obtain 
Eq.\eqref{eq:inver}. The initial conditions, i.e. $r_{0}, r_{1},\ldots, 
r_{S-1}$, are the same as in the first derivation and this leads to the 
conclusion that the following equation holds:
\begin{equation}
-r_{k}=R_{k}=\frac{\mathcal{H}_{k}}{\mathcal{H}_{k-1}}\frac{1}{\gamma_{k}^{-1}}=-\frac{a_{k+1}}{a_{k}}.
\label{eq:ha}
\end{equation}

As is well known, the Hill determinant method requires that for the quasinormal modes \cite{hill}
\begin{equation}
 \lim_{k \to \infty}\mathcal{H}_{k}=0.
\end{equation}
This demonstrates the equivalence of the Hill determinant method and our 
original approach based on Eq.~\eqref{eq:a}, as suggested by the truncation 
process of the Hill determinant \eqref{eq:mat}, where we seek the values 
$\omega$ for which non trivial solutions of the truncated system \eqref{eq:recd} 
exist, such that, effectively, $a_{k+1}=0$.

A standard way to obtain reasonable results is to set $n$-th Hill determinant to zero
for a sufficiently large $n$ and identify solutions representing approximate
quasinormal mode frequencies. An improvement of this method is to construct a
sequence of equations  $\mathcal{H}_{n}=0$ and search for solutions that, while
migrating in the complex plane, remain bound. For such successive approximations
of the sought frequency, convergence acceleration techniques should be applied
to improve the results.

Now, in a close analogy to Nollert's approach~\cite{nollert} and its extension 
propounded by Zhidenko~\cite{zhidenko}, instead of equating either representation
of the main equation~\eqref{eq:ha} to zero, we shall approximate it with its 
asymptotic expansion\footnote{To be precise, the notion of the remainder was
introduced to the black hole physics by Leaver in his original work~\cite{leaver}, see also
the discussion in the
appendix of Ref.~\cite{leavergauss}}. It is done with an ansatz
\begin{equation}
R_{N}\sim \widetilde{R}_{N} = \sum^{\infty}_{i=0} \frac{c_{i}}{N^{i/p}}\label{eq:nollert}
\end{equation}
as $N \to \infty,$ where $p=2$.
In general, we may consider other expansions of this type with a different 
choice of $p$~\cite{res, benors}. An inappropriate choice of
$p$, for example $p=1$ in our case,  would lead to the singular 
coefficients $c$ and therefore would be easily detectable.

To calculate the coefficients of the asymptotic expansion, we
insert~\eqref{eq:nollert} into Eq.~\eqref{eq:remainderrec}, collect the terms 
with the like powers of $N,$ and solve the thus obtained system of equations of 
ascending complexity. We have to choose $c_{0}$ and $c_{1}$ to satisfy the asymptotic 
expansion~\eqref{eq:aa}, representing the solution which is convergent at spatial 
infinity with a unit radius of convergence, whereas the higher order terms $c$ 
are defined uniquely. Finally, our numerical procedure for calculating the
frequencies of the quasinormal modes requires solving of the following equation:
\begin{equation}
 \mathcal{H}_{L}-\mathcal{H}_{L-1} \gamma_{L}^{-1} \widetilde{R}_{L}=0.\label{eq:solve}
\end{equation}
This procedure can be viewed as taking into account asymptotic behavior of the Hill
determinants rather then requiring their vanishing.

Of course, it is equally straightforward to incorporate the term describing 
the remainder into the two other approaches we have previously analyzed. First, 
consider the downward recurrence \eqref{eq:inver}. Substituting the right-hand 
side of equation \eqref{eq:down-up} with the expression for the remainder gives
\begin{equation}
 \sum_{i=0}^{S} \gamma^{i}_{L}\prod^{i}_{j=1}\frac{1}{r_{L-j}} = \gamma^{-1}_{L} \widetilde{R}_{L}.
 \label{eq:frac-with-tail}
\end{equation} 
The asymptotic form of the remainder is also easy to include in 
the approach we have discussed initially. We cannot simply introduce the asymptotic 
for $a_{L}$ in our original approach based on Eq.~\eqref{eq:a}, as it contains 
some unspecified constant. This is why we should analyze the ratio $a_{L}/a_{L-1}$ instead.

It should be noted that performing Gaussian elimination \cite{zhidenko} on 
\eqref{eq:recd} is equivalent to the analogous operation on the Hill determinant. This is 
possible due to the form of the initial conditions \eqref{eq:reci}. The resulting 
three-term recurrence relation leaves only two independent solutions, and because of 
the initial conditions, these are the ones represented by the asymptotic behaviors 
corresponding to the unit radius of convergence. Nevertheless, due to the invariance 
of the determinant under the action of the Gaussian elimination\footnote{In our problem 
we will only add a multiple of one row to another row~\cite{zhidenko}},
the resulting equation we obtain
by applying the Pincherle theorem \cite{gautschi, leaver} and making use of 
\eqref{eq:nollert} would lead to the same equation \eqref{eq:solve}.

To construct the approximate  values of the quasinormal modes (in contrast to 
the very precise ones), we propose numerically solving either 
Eq.~\eqref{eq:solve} or Eq.~\eqref{eq:frac-with-tail} with a small order (size) 
of the Hill matrix, say $L=50$, and  with the first few Nollert terms, for 
example up to $c_{4}$. Apart from enhancing the accuracy of the results - which 
is very poor for small $L$, particularly for higher overtones - the tail 
approximation also excludes results that can be regarded as numerical artifacts 
rather than approximations of quasinormal modes. However, there is still one 
nonphysical solution: it is unstable in the sense that its value changes with 
$L$. It is purely imaginary and appears to converge to zero as $L$ increases, 
making it easily distinguishable from the physical solutions.  Then, to obtain very accurate 
results, we increase both $L$ and the number of terms in the asymptotic 
expansion of the remainder. Finally, we search for a specific root representing the 
complex frequency of the quasinormal mode, using an initial guess obtained from 
the low-precision calculations described above.

The consecutive approximations of a quasinormal mode, obtained by solving (21) 
with different $L$, inherit their spiraling convergence from those without the 
remainder approximation, as demonstrated in \cite{matyjasek2021accurate, 
analog}. Finally, we can apply the Wynn acceleration algorithm to obtain 
arguably the most precise results available in the literature, with relatively 
short computation times.

The typical time scales of the computations on a modern home computer are as 
follows\footnote{Codes with our implementation will be shared on request. All 
the calculations were performed using Wolfram Language, developed by Wolfram 
Research, Inc.}: The initial approximate values are calculated in a few minutes, 
the first few (5 to 10) Nollert terms are calculated in seconds, although 
higher-order terms, especially for higher-order recurrences, become increasingly 
challenging. The computation of the most complicated terms takes a few 
hours, however, it should be remembered that such calculations for a specific 
black hole dimension have to be performed only once. The calculation of all the 
equations simultaneously for consecutive Hill determinants of the matrices which 
size is up to $1000\times1000$ are calculated in seconds using 
\eqref{eq:hillrec}. Finding the approximants of the specific quasinormal mode 
for all the Hill determinant equations, $\mathcal{H}_{k} =0$ with  $k\leq 500$ 
takes around 5 minutes. Finally, applying the Wynn acceleration to the 
approximants takes only a few seconds. Whenever possible, we tried to perform 
our calculations in analytical form, resorting to numerical methods only in the 
final stage. The expected cost of the calculations, in terms of the time they 
take, is balanced by the stability, reliability, and quality of the results. Of 
course, in most cases, it is possible to use numerical methods from the very 
beginning. The purely numerical approach can be particularly advantageous for 
extremely high overtones~\cite{high, gauss}.

\section{Illustrative numerical results}
\label{sec:secIV}

We shall illustrate our analysis with the three interesting examples. We start 
with the static and spherically symmetric five- and seven-dimensional black 
holes, which, although interesting in their own right, exhibit a similar 
structure of equations describing radial perturbations to those of $(2+1)$- and 
$(3+1)$-dimensional acoustic black holes~\cite{analog}. Indeed, the equation 
describing the radial perturbations of the static $(2+1)$-dimensional acoustic 
black hole is a special case of the master equation \eqref{eq:master1} with 
$D = 5$, $j = 2/3$, and $l = m - 1$ for $m \geq 1$. Similarly, taking $D=7$, $j=3/5$, 
and $l=(2 s-3)/2$ yields the radial equation for the $(3+1)$-dimensional 
acoustic black hole. Additionally, we analyze the six-dimensional 
Schwarzschild-Tangherlini black hole. The specific cases we have chosen 
should be representative of problems 
related to determining the quasinormal modes of Schwarzschild-Tangherlini 
black holes, with dimension not exceeding $D =9$. They should also be 
representative of other configurations for which the modes can be determined 
using recursive methods.

\subsection{The general five-dimensional case and $(2+1)$-dimensional acoustic black holes}

To begin with, let us consider the general five-dimensional case.  Substituting 
\eqref{eq:expansion_odd} into the master equation yields a four-term recurrence 
relation
\begin{equation}
 \gm{k}{-1} a_{k+1} + \gm{k}{0} a_{k} + \gm{k}{1} a_{k-1} + \gm{k}{2} a_{k-2} =0,
                    \label{recurrence_1}
\end{equation}
%----
where the recurrence coefficients are given by
%----
\begin{equation}
\begin{array}{lcl}
\gm{k}{-1} &=& -8 (1+k)(1+k + \rho),\\
\gm{k}{0} &=& 20 k^2+4 k (8 \rho +5)+4 l^2+8 l+16 \rho ^2+16 \rho +9 (1-j^2) +3,\\
\gm{k}{1} &=& -2 \left[8 k^2+8 k \rho +9 (1-j^{2}) -8\right],\\
\gm{k}{2} &=& 4 k^2-4 k+9 (1-j^2)-8
 \end{array}
 \label{rec1}
\end{equation}
and $\rho=-i\omega$.
Similarly, specializing \eqref{eq:hillrec} to the five-dimensional case, we are left with 
\begin{equation}
 \mathcal{H}_{k}=\gm{k}{0}\, \mathcal{H}_{k-1}-\gm{k}{1} \gm{k-1}{-1} \mathcal{H}_{k-2} +
 \gm{k}{2} \gm{k-1}{-1} \gm{k-2}{-1}  \mathcal{H}_{k-3}.
\end{equation}
And finally,  the recurrence for the remainder obtained from \eqref{eq:remainderrec} is given by
\begin{equation}
R_{k}\left( \gm{k+2}{-1} R_{k+2} R_{k+1}-\gm{k+2}{0} R_{k+1}+\gm{k+2}{1}\right)-\gm{k+2}{2} = 0.
\end{equation}

\begin{figure}[htb]
    \centering
    \includegraphics[width=0.7 \textwidth, clip]{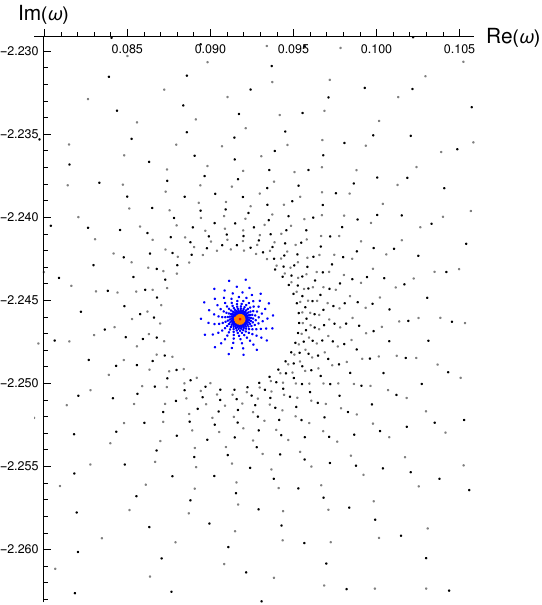}
    \caption{Approximants to the $m=1$, $n=2$ mode of the $(2+1)$ hydrodynamic black hole 
    (its value is $\approx0.091778997 - 2.246129591 i$)
    emerging from $100\times100$ to $500\times500$ Hill determinants. Black dots represent 
    approximants without tail approximation; Gray dots \texttwelveudash approximants with 
    tail up to the $c_{0}$ term (We see that this does not effect the convergence much, 
    which is understandable as both the correct and incorrect asymptotics are then identical); 
    Blue dots \texttwelveudash approximants with tail up to the $c_{1}$ term;
    Orange spot represent concentration of approximants with terms up to the $c_{2}$; 
    Red dot represent the value calculated using Wynn acceleration algorithm of either 
    group of dots.}
    \label{fig:1}
\end{figure}

\begin{figure}[htb]
    \centering
    \includegraphics[width=0.7 \textwidth, clip]{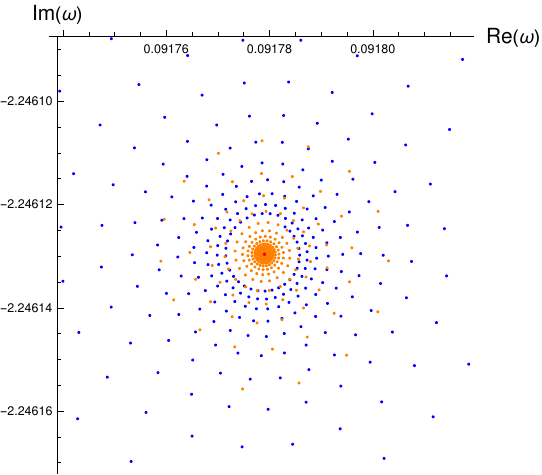}
    \caption{Zoomed in region in the vicinity of the red dot in figure \ref{fig:1}.}
    \label{fig:2}
\end{figure}

\begin{figure}[htb]
    \centering
    \includegraphics[width=0.7 \textwidth, clip]{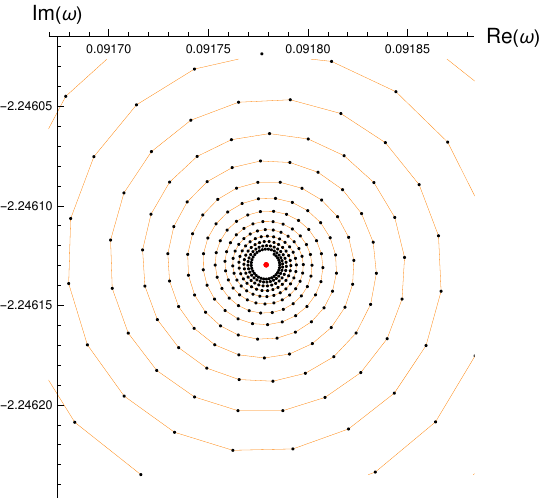}
    \caption{Further zoom towards the red dot, only group of root 
    approximants emerging from the Nollert sum up to $c_{2}$ are shown. Consecutive 
    convergents are connected with lines to underline their vortex behavior.}
    \label{fig:3}
\end{figure}

\begin{figure}[htb]
    \centering
    \includegraphics[width=0.7 \textwidth, clip]{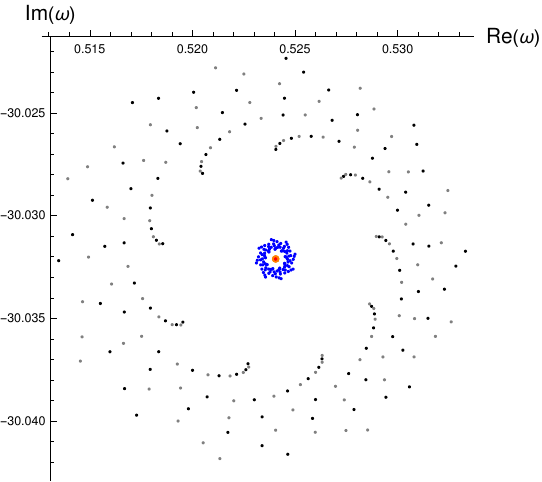}
    \caption{Approximants to the $l=4$, $n=20$ quasinormal mode of the 
electromagnetic vector perturbations ($j=1/2$) of 6-dimensional 
Schwarzschild-Tangherlini black hole. Details and legend for the figure are the 
same as for Figure~\ref{fig:1}, with the exception that only approximants from 
$400\times400$ to $500\times500$ determinants are shown as including all the 
lower order ones would lead to overlap between the different groups.}
    \label{fig:4}
\end{figure}

Now, we demonstrate how the roots of the Hill determinants, identified as 
consecutive approximations of the frequency of a given mode, migrate in the 
complex plane. In this subsection we shall consider only  the  $m=1$, $n=2$ mode 
of the $(2+1)$-dimensional hydrodynamic black hole, which satisfies the master equation
in five dimensions. In choosing this particular 
mode, we have two goals to achieve: first, we want to illustrate our approach with 
the mode which is neither, because of the rapid convergence, too easy to be 
calculated, nor the one that causes any extraordinary difficulties. Our second 
reason is to go beyond the values of $j$ (and implicitly $l$) given in~\eqref{eq:typyzab}.

In Fig.~\ref{fig:1}, a few groups of approximants are 
plotted. The outermost group represents approximants calculated from the Hill 
determinants either without applying any convergence acceleration or with the 
$c_{0}$ term retained.  (The first few coefficients $c_{i}$ are given in Appendix). 
A deeper internal structure represents solutions
constructed by taking the first two tail terms into account. Finally, the small 
spot at the center represents solutions obtained with the first three terms. In 
Figs.~\ref{fig:2} and \ref{fig:3}, the enlarged regions of each group are shown. 
We see that by adding successive terms of the Nollert series, we obtain a 
qualitative improvement in the result. This holds at least for the first few 
terms. 

For the fundamental and  low-lying modes, the accuracy we obtain  from a 
sufficiently large Hill matrices without any convergence acceleration is quite 
impressive. On the other hand, for the more challenging modes, the procedure is 
more complicated. In this case, if we are not satisfied with the accuracy of the 
answer, some form of the convergence acceleration is necessary\footnote{For the 
next higher lying mode (see section V), the possibility of stable convergence 
itself without any acceleration, is questionable.}. We see that adding 
consecutive $c_{i}$ terms to the Nollert sum moves the convergents towards the 
assumed limit in agreement with the value we obtain from the Wynn acceleration 
algorithm applied to the original sequence, and, therefore, both methods, in a sense, confirm 
one another.

Let us compare the results obtained with different truncations of Nollert’s sum 
in more detail. Our discussion will be greatly simplified if we define $\omega_{d}$ 
as the 
Wynn-accelerated value of the series of approximants for the quasinormal mode 
frequency, calculated using Hill determinants for matrices ranging from 
$100\times100$ to $500 \times 500$ and taking into account the Nollert tail 
terms up to $c_{d}$. We take $\omega_{30}$, as the reference value  with which 
we will compare all other values. For a frequency $\Omega$, that we get from  
$500\times500$ Hill determinant, the difference between it and $\omega_{30}$ is 
$\Delta \Omega \approx0.002+0.003i$. The difference between Wynn's acceleration 
value of the Hill determinants $\omega$ and $\omega_{30}$ is $\Delta \omega 
\sim10^{-13}+ 10^{-14}i$. Carrying on we get differences between values with 
different remainder truncation: 
$\Delta \omega_{0}\sim10^{-13}+ 10^{-12}i$,
$\Delta \omega_{1}\sim10^{-36}+ 10^{-36}i$,
$\Delta \omega_{2}\sim10^{-47}+ 10^{-47}i$,
$\Delta \omega_{10}\sim10^{-78}+ 10^{-78}i$,
$\Delta \omega_{20} \sim10^{-93}+ 10^{-93}i$
and
$\Delta \omega_{29} \sim10^{-102}+ 10^{-102}i$.

We have also calculated the Wynn-accelerated value of the frequency approximants 
for the determinants of the matrices from $100\times100$ to $1500\times1500$ 
without tail approximation $\omega^{100-1500}$ to confirm that they move towards 
the assumed limit in the closest vicinity of $\omega_{30}$: $\Delta 
\omega^{100-1500} \sim 10^{-44}+ 10^{-45}i$. If we take the Wynn acceleration of 
those determinants with the tail approximation up to 30-th term, the difference 
between them and $\omega_{30}$ is $\sim 10^{-104}+ 10^{-104}i$, which suggest 
that $\omega_{30}$ gives the correct result up to about 100 digits.

Finally, let us also compare  $\omega_{30}$ with the results we would obtain with $30$ Nollert 
tail terms applied to the $L$-th determinant without Wynn acceleration. We will 
denote the difference between such result and $\omega_{30}$ as $\Delta 
\omega^{L}$. $\Delta \omega^{500}\sim10^{-29}+ 10^{-29}i$;
$\Delta \omega^{1000}\sim10^{-35}+ 10^{-35}i$;
$\Delta \omega^{1500}\sim10^{-38}+ 10^{-38}i$;
$\Delta \omega^{2000}\sim10^{-40}+ 10^{-40}i$;
$\Delta \omega^{2500} \sim10^{-42}+ 10^{-42}i$. 
We can see that the accuracy in this approach grows quite slowly.

\subsection{The general seven-dimensional case and $(3+1)$-dimensional acoustic black holes}

Following the steps of our previous demonstration, we have calculated the first 
few fundamental modes and their overtones for the massless scalar (gravitational tensor) 
perturbations $(j = 0)$ 
of the seven-dimensional Schwarzschild-Tangherlini black hole and have arranged 
the results in Table~\ref{tab1}.  We have used the solution of the master equation in the 
form proposed in Ref. \cite{analog} 
\begin{equation}
 \psi(r) = \left(\frac{r-1}{r+1}\right)^{-i \omega/4} e^{i \omega r}
e^{-i \omega \arctan( r)/2} \sum_{k=0}^{\infty} a_{k} \left(\frac{r-1}{r} \right)^{k}, \label{eq:arc}
\end{equation}
which leads to the six-term recurrence instead of the eight-term one 
\begin{equation}
\begin{array}{lcl}
\gm{k}{-1} &=&  -8 (k+1) (2 k+\rho +2)  ,\\ 
\gm{k}{0} &=& 56 k^2+8 k (6 \rho +7)+4 l (l+4)+12 \rho ^2+24 \rho +25 (1-j^{2})+15,\\
\gm{k}{1} &=& \left[20 k^2+14 k \rho +2 \rho ^2+25 (1-j^{2})-20\right],\\
\gm{k}{2} &=&  2 \left[30 k^2+2 k (8 \rho -15)+2 \rho ^2-8 \rho +75 (1-j^{2})-60\right] ,\\
\gm{k}{3} &=& -24 k^2-8 k (\rho -6)+8 (\rho +9)-100 (1-j^{2}),\\
\gm{k}{4} &=&4 k^2-12 k+25 (1-j^{2} )-16.
 \end{array}
 \label{rec1}
\end{equation}
For each $0\leq l \leq 3$, we have calculated the fundamental mode and its first two overtones. 
Additionally, to check the computational stability in the case when $\Re(\omega)< |\Im(\omega)|$,
we have also calculated 9-th overtones.

\begin{table}
 \caption{Massless scalar (gravitational tensor) quasinormal modes of the 
7-dimensional Schwarzschild-Tangherlini black hole. $l$ is the multipole number, 
$n$ the overtone number, $\omega$ is the complex frequency resulting from Wynn 
acceleration of Hill determinants from $100\times 100$ to $500\times 500$ with 
Nollert tail terms up to $c_{15}$. The results of $\omega$ are rounded to 20 
digits.  The lack of implementation of the Wynn acceleration for $n=9$ modes 
would lead to correct result up to about 10 decimal places.}
\begin{ruledtabular}
\begin{tabular}{ccccc}
$ l$ & $n$ & $\omega$  \\
\colrule
$0$&$  0$&$ 1.2705405420674844936 - 0.6657777484940844370 i$\\
$ $&$1 $&$ 0.6834280550829221450 - 2.4387924655957958174 i
$\\
$ $& $  2$&$ 0.5042084172777775591 - 4.6369542021672305536 i$\\
$ $& $  9$&$           0.3918741425502854957 - 18.8780814844864886150 i
$\\
$1$&$  0$&$ 1.8813962898067009330 - 0.6410765449925537831 i

$\\
$ $&$  1$&$ 1.3926588635644419050 - 2.0657071981696225173 i

$\\
$ $&$  2$&$ 0.7636568558659216879 - 4.1765987015543314493 i
$\\
$ $&$  9$&$ 0.4400521954241858894 - 18.7387484682812457866 i
$\\

$2$&$  0$&$ 2.4967803477320673916 - 0.6318819213285035104 i

$\\
$ $&$  1$&$ 2.1372690423629634552 - 1.9611691867830599187 i

$\\
$ $&$  2$&$ 1.4011726864592502091 - 3.6414741303614181889 i
$
\\
$ $&$  9$&$ 0.5107405227778542691 - 18.5393990613782136359 i

$
\\
$3$&$  0$&$ 3.1140725371342787624 - 0.6276147383334851258 i

$\\
$ $&$  1$&$2.8306306034610711898 - 1.9213975318608638971 i

$\\
$ $&$  2$&$ 2.2293221775716185709 - 3.3836727804017218864 i

$\\
$ $&$  9$&$ 0.6113430265063098204 - 18.2762911255452473162 i
$\\

\end{tabular}
\label{tab1}
\end{ruledtabular}
\end{table}

\begin{table}
 \caption{Quasinormal mode spectrum of the electromagnetic vector perturbations 
($j=1/2$) of the six-dimensional Schwarzschild-Tangherlini black hole. The computational 
strategy is identical to the one adopted in the calculations of the modes listed in Table~\ref{tab1}.}
\begin{ruledtabular}
\begin{tabular}{ccccc}
$ l$ & $n$ & $\omega$  \\
\colrule

$1$&$  0$&$ 1.3999400314954935394 - 0.4982260948677470476 i
$\\
$ $&$  1$&$ 1.0951781787045935667 - 1.6061082680722314566 i
$\\
$ $&$  2$&$ 0.7018823450043227931 - 3.0553437787592727320 i
$\\
$ $&$  17$&$ 0.131018471487378262 - 26.020751665520504764 i
$\\

$2$&$  0$&$ 1.9783041247752266468 - 0.4964229788580986953 i

$\\
$ $&$  1$&$ 1.7525021749735041514 - 1.5419339117617781247 i
$\\
$ $&$  2$&$ 1.3439355956672251518 - 2.7827675389874555525 i
$
\\
$ $&$  18$&$ 0.246864908655348317 - 27.368648608574783871 i
$
\\
$3$&$  0$&$ 2.5533344325272784565 - 0.4956300618599753434 i

$\\
$ $&$  1$&$2.3764027060977187628 - 1.5174554256492204046 i

$\\
$ $&$  2$&$ 2.0313017726197248102 - 2.6480664431633192274 i
$\\
$ $&$  17$&$ 0.411239587920173643 - 25.670392308423745917 i
$\\

$4$&$  0$&$ 3.1268028274221800720 - 0.4952057066457678160 i
$\\
$ $&$  1$&$2.9817220300113130424 - 1.5056198133863101051 i
$\\
$ $&$  2$&$ 2.6935008768148631664 - 2.5839936316504141436 i

$\\
$ $&$  20$&$ 0.524052589363473999 - 30.032105024086907721 i
$\\

\end{tabular}
\label{tab2}
\end{ruledtabular}
\end{table}

\subsection{The six-dimensional case}

Finally, for the six-dimensional Schwarzschild-Tangherlini black holes, we have 
calculated the quasinormal mode frequencies of the electromagnetic vector 
perturbations ($j = 1/2$) and present the results in Tab.~\ref{tab2}. This 
problem is related to a five-term recurrence,
\begin{equation}
 \sum_{s = -1}^{3} \gm{k}{s} a_{k} = 0,
\end{equation}
where the  coefficients have the form:
\begin{equation}
\begin{array}{lcl}
\gm{k}{-1} &=& (1 + k) (3 + 3 k + 2 \rho),\\
\gm{k}{0} &=& -6 + 4j^2 - 9k^2 - 3l - l^2 - 6\rho - 4\rho^2 - 3k(3 + 4\rho),\\
\gm{k}{1} &=& -2 \left[-5 + 6(1 - j^2) + 5k^2 + 6k\rho + 2\rho^2
\right],\\
\gm{k}{2} &=& -2 + 12j^2 + 5k - 5k^2 + 2\rho - 4k\rho
,\\
\gm{k}{3} &=& -4j^2 + (-1 + k)^2.
 \end{array}
 \label{rec11}
\end{equation}
For each $1 \leq l \leq 4$,  we have calculated the fundamental modes and their 
first two overtones. Additionally, we have computed one higher overtone as a 
representative example for the more complicated cases. For example, for the $l=4$ 
mode, we have chosen $n=20$ and illustrated the migration of the approximants on 
the complex plane in Figure \ref{fig:4}. An examination of Table \ref{tab2} 
reveals that this mode has a relatively large real part, allowing for stable 
calculations even without tail implementation, in contrast to the cases for 
$l=1,2$. Nevertheless, the convergence remains poor in the absence of acceleration. 
A computational strategy relying solely on the first 16 terms of the Nollert 
expansion (without employing the Wynn algorithm) would yield the results with 
approximately 15 correct digits. Once again, we see that 
a well-considered application of both the Nollert terms and the Wynn convergence 
acceleration is essential for obtaining highly precise results.

\section{Conclusions.}
\label{sec:secV}

We have developed and numerically tested certain extensions  of the Leaver 
techniques. Our analysis provides greater flexibility in implementations and 
demonstrates, for instance, that the Hill determinant and continued fraction 
methods are merely different manifestations of the same underlying approach. 
This correspondence enables (with the help of Eq. \eqref{eq:ha}) the application 
of the asymptotic expansion of the remainder in the Hill determinant method and 
makes the  Gaussian elimination unnecessary. We have also derived a few useful 
recurrence formulas and demonstrated - with a high degree of certainty - that 
the consecutive approximants of the quasinormal modes emerging from the Hill 
determinants demonstrate regular spiraling pattern even if the tail terms are 
taken into account, which is a strong argument for using the convergence 
acceleration techniques, with the Wynn acceleration being our first choice. In 
particular, the Wynn acceleration and the improvements arising solely from the 
tail implementation validate each other's results.

Let us conclude with some remarks regarding possible difficulties with the 
implementation. We previously stated that for considered problems there are two 
independent asymptotics of the form
\begin{equation}
 a_{k+1}/a_{k} \sim 1\pm\frac{\sqrt{2 \rho}}{\sqrt{k}}+\ldots.\label{eq:+-}
\end{equation}
If the real part of the second term in Eq. \eqref{eq:+-} is non-zero, one of 
those asymptotics (with the minus sign if the principal branch of square root is 
considered) represent a convergent solution at infinity and 
the other (with the plus sign) represents solution which is divergent at infinity. 
For the purely imaginary values this logic does not apply as the real 
part of the second term is zero. The applicability of the recurrence based 
methods then require additional consideration. For the critique of the continued 
fraction method (and consequently other methods based on the recurrences) in 
this and similar contexts, consult \cite{purely, exact}.

When searching for a specific root, our calculations suggest that the modes with 
a relatively small real part cannot be  calculated in a stable manner without 
the tail. A good example is the $m=1$, $n=3$ mode of the $(2+1)$ hydrodynamic 
black hole, with its value being approximately $0.0349613 - 3.25971 i$. The main 
issue may be illustrated by the fact that many numerically calculated 
convergents are located  on the imaginary axis. Interestingly, applying the Wynn 
acceleration to the consecutive convergents from the Hill determinants still 
yields a few correct digits of the mode, probably due to the stable behavior of 
the convergents further away from the imaginary axis. This problem is resolved 
by implementing the remainder asymptotic \cite{nollert, zhidenko}. We should 
keep in mind that if the remainder contains too many terms compared to a 
relatively small Hill determinant, new nonphysical roots may appear. Therefore, 
special care is needed, as the role of the remainder is to improve the accuracy 
of already existing solutions, not to generate new ones. Fortunately, according 
to our practice, they typically do not pose any problems in the calculation of 
quasinormal modes, as they do not appear in the strategies proposed by us in the 
article.

We conclude with one remark.  In this work, we present the results of our 
calculations with an accuracy of approximately 19 decimal places. We believe 
that all these results are correct. The reason for such high accuracy is (at 
least) twofold: First, we aim to conduct a thorough analysis of various features 
of the proposed methods, such as their generality, the exactness of the results, 
and overall performance. Secondly, we want to provide the high-quality results 
that can serve as a comparative material for other methods. Moreover, having the 
exact numerical results allows one to test various hypotheses regarding the 
global nature of the quasinormal modes.

\begin{acknowledgments}
J. M. was partially supported by Grant No. 2022/45/B/ST2/00013 
of the National Science Center, Poland.\end{acknowledgments}
%%%%%%%%%%%%%%%%%%%%%%% 
%\clearpage
%%%%%%%%%%%%%%%%%%%%%%%

\appendix*
\section{The coefficients of the asymptotic expansion}

Let us discuss the structure of $c_{i}$ coefficients in \eqref{eq:nollert}. The
value of $c_{0}$ is $-1$, which can easily be deduced 
from the radius of convergence of the series.
It turns out that the first three coefficients are the same for 
all the configurations studied in this article 
with\footnote{We have also calculated the Nollert terms for the Dirac
quasinormal modes of the Schwarzschild black hole~\cite{dirac} and found that $c_{0}$ and $c_{1}$ coefficients
are given by the same values as in our case.}
\begin{equation}
c_{0} = -1,
\end{equation}
\begin{equation}
c_{1} = \sqrt{2} \sqrt{\rho},
\end{equation}
\begin{equation}
c_{2} = \frac{3}{4}-\rho.
\end{equation}
The remaining ones can be calculated for a general perturbation type $j$ and
multipole number $l$.  It should be noted that $l$ appears already in the 
$c_{3}$ coefficient, whereas $j$ appears only, depending on the dimension of the 
black hole, in the  higher terms. Indeed, for $D=5, \,6$ and $7$, the parameter $j$ 
appears for the first time in $c_{5}$, $c_{6}$ and $c_{7}$, respectively. 

Below, we  list the expansion coefficients  truncating it at the first 
coefficient in which the dependence on the type of perturbation appears. 
It should be noted that we are not confined to the values of the parameter $j$ 
as given in \eqref{eq:typyzab}. 

For the five-dimensional Schwarzschild-Tangherlini black hole, one has
\begin{equation}
c_{3} = \frac{15 + 32l + 16l^2 - 64\rho + 16\rho^2}{32 \sqrt{2}\sqrt{\rho}},
\end{equation}
\begin{equation}
c_{4} = -\frac{15 + 32l + 16l^2 + 96\rho - 144 \rho^2}{128 \rho},
\end{equation}
\begin{eqnarray}
c_{5} &=& \frac{1}{4096 \sqrt{2} \rho^{3/2}}\left[ 135 - 1024l^3 - 256l^4 - 1920\rho - 32(-533 + 288j^2)\rho^2-2048\rho^3   \right. \nonumber \\
&& \left.
  - 256\rho^4 - 64l(3 + 64\rho + 16\rho^2) - 32l^2(35 + 64\rho + 16\rho^2\right].
\end{eqnarray}

A similar calculation for the $D=6$ black hole gives
\begin{equation}
c_{3} = \frac{35 + 48l + 16l^2 - 64\rho + 16\rho^2}{32 \sqrt{2}\sqrt{\rho}},
\end{equation}
\begin{equation}
c_{4} = -\frac{35 + 48l + 16l^2 + 96\rho - 144\rho^2}{128 \rho},
\end{equation}
\begin{eqnarray}
c_{5} &=& \frac{1}{4096 \sqrt{2} \rho^{3/2}}\left[ -385 - 1536l^3 - 256l^4 - 4480\rho + 7200\rho^2 - 2048\rho^3
\right. \nonumber \\
&& \left.
- 256\rho^4 - 96l(23 + 64\rho + 16\rho^2) - 32l^2(95 + 64\rho + 16\rho^2\right],
\end{eqnarray}
\begin{eqnarray}
c_{6} &=&\frac{1}{4096 \rho^2}\left[
805 + 1536l^3 + 256l^4 + 1120\rho + 3072\rho^2 + 512(-41 + 32j^2)\rho^3
\right.\nonumber \\
 &&\left.
 - 256\rho^4 + 96l(29 + 16\rho) + 32l^2(101 + 16\rho) \right].
\end{eqnarray}

Finally, take $D=7$.  The computed coefficients assume the form:
\begin{equation}
c_{3} = \frac{16 l^2+64 l-16 \rho^2-64 \rho+63}{32 \sqrt{2}\sqrt{\rho}},
\end{equation}
\begin{equation}
c_{4} = -\frac{16 l^2+64 l-48 \rho^2+96\rho+63}{128 \rho},
\end{equation}
\begin{eqnarray}
c_{5} &=& \frac{1}{4096 \sqrt{2} \rho^{3/2}}\left[ -256 l^4-2048 l^3-32 l^2 \left(112 \rho^2+64\rho+179\right) \right. \nonumber \\
&& \left. 
-128 l \left(112 \rho^2+64\rho+51\right)+12032 \rho^4-2048 \rho^3-5792\rho^2-8064 \rho-2457\right],
\end{eqnarray}
\begin{eqnarray}
c_{6} &=&\frac{1}{4096 \rho^2}\left[ 
32 l^2 \left(-128 \rho^3+32 \rho^2+16\rho+185\right) -128 l \left(128 \rho^3-32 \rho^2-16\rho-57\right)\right.\nonumber \\
 &&\left.
 +12288 \rho^5+23296 \rho^4-19968\rho^3+7104 \rho^2+2016 \rho + 3213+ 256 l^4+2048 l^3 \right],
\end{eqnarray}
 \begin{eqnarray}
c_{7} &=& \frac{1}{262144\sqrt{2} \rho^{5/2}}\left\{
-256 \left(12800 j^2-32589\right) \rho^4
+4096 l^6+49152l^5 + 256 l^4 \left(208 \rho^2+64 \rho+845\right)\right.
\nonumber \\
&&
+2048 l^3\left(208 \rho^2+64 \rho+205\right)
+
16 l^2 \left[41728\rho^4+2048 \rho^3+60256 \rho^2+112 \left(208\rho^2+64 \rho+77\right)\right.
\nonumber \\
&& \left.
+15744 \rho+9843\right]+64 l \left(41728 \rho^4+2048 \rho^3+30304 \rho^2+6528\rho-1245\right)
-1904640 \rho^6 \nonumber \\
&&\left.
+ 5193728 \rho^5-403456\rho^3+1083600 \rho^2+157248 \rho-134001 \right\}.
\end{eqnarray}
 We conclude with the remark that one might consider the possibility of calculating these
coefficients recursively, using a procedure similar to the one presented in Ref. [32].\\

%\addcontentsline{toc}{section}{References}
%\bibliography{kb}
%\bibliographystyle{ieeetr}

\bibliography{kb}

\end{document}